\documentclass[preprint,aps]{revtex4}
\usepackage{epsfig,psfig}
\begin{document}
\title{Do two flavour oscillations explain both KamLAND Data and 
the Solar Neutrino Spectrum?}
\author
{Bipin Singh Koranga, Mohan Narayan and S. Uma Sankar}
\affiliation
{Department of Physics, Indian Institute of Technology, 
Powai, Mumbai 400 076}
\date{\today}

\begin{abstract}
The recent measurement of $\Delta_{sol}$ by the KamLAND experiment
with very small errors, makes definitive predictions for the
energy dependence of the solar neutrino survival probability
$P_{ee}$. We fix $\Delta_{sol}$ to be the KamLAND best fit
value of $8 \times 10^{-5}$ eV$^2$ and study the energy dependence
of $P_{ee}$ for solar neutrinos, in the framework of two flavour 
oscillations and also of three flavour oscillations. For the case of
two flavour oscillations, $P_{ee}$ has a measurable slope in the
$5-8$ MeV range but the solar spectrum measurements in this range
find $P_{ee}$ to be flat. The predicted values of $P_{ee}$, even for
the best fit value of $\theta_{sol}$, differ by $2 \ {\rm to} \ 3 
\ \sigma$ from the Super-K measured values in each of the three energy 
bins of the $5-8$ MeV range. If future measurements of solar neutrinos 
by Super-K and SNO find a flat spectrum with reduced error bars (by a 
factor of $2$), it will imply that two flavour oscillations can no longer 
explain both KamLAND data and the solar spectrum. However a flat solar 
neutrino spectrum and the $\Delta_{sol}$ measured by KamLAND can
be reconciled in a three flavour oscillation framework with a moderate 
value of $\theta_{13} \simeq 13^\circ$.
\end{abstract}

\maketitle

\section{Introduction}

Recently the KamLAND experiment has measured $\Delta_{sol}$ with great 
accuracy \cite{kamlandsp}. Their fit to the data, assuming neutrino 
oscillations, gives the result:
\begin{equation}
\Delta_{sol} = (8.0 \pm 0.5) \times 10^{-5} {\rm eV}^2.
\end{equation}
The error in this measurement, already as low as $6 \%$, can be further 
improved with more data. The accuracy of KamLAND in determining the mixing 
angle $\theta_{sol}$ is not so good. The reason, for this great disparity in 
the accuracies of the two oscillation parameters, is the following: 
The anti-neutrino survival probability, for two flavour oscillations, 
is given by 
\begin{equation}
P_{\bar{e} \bar{e}} = 1 - \sin^2 2 \theta_{sol} 
\sin^2 \left( 1.27 \Delta_{sol} L / E \right).
\end{equation} 
This expression has minima when the energy takes the following values,
$E_p = 2.54 \Delta_{sol} L/(2p+1)\pi$,
where $p$ is an integer. Thus the position of the minimum is
dictated by $\Delta_{sol}$ and the depth of the minimum is 
dictated by $\theta_{sol}$. The inverse beta-decay spectrum of positrons, 
measured by KamLAND as a function of the anti-neutrino energy, is 
a product of the spectrum in the case of no oscillations (which 
is a smooth function) and the above survival probability. The 
minima of the survival probability lead to kinks (or distortions) 
in the measured spectrum. Given the baseline of KamLAND 
and the energy range of the anti-neutrinos only the minimum 
corresponding to $p=1$ is relevant. The position of the kink 
produced by this minimum can be measured with great accuracy 
which leads to $\Delta_{sol}$ being determined with great accuracy. 
The amount of distortion in the kink can be measured only with a limited 
accuracy which means that $\theta_{sol}$ can be determined only with a
limited accuracy. Detailed stastical calculations also lead to the 
same conclusion \cite{srubabatikaml}.

Given the accurate measurement of $\Delta_{sol}$ by KamLAND, it is
worth raising the question: What is its impact on the solar neutrino
problem? In the $\Delta_{sol} - \theta_{sol}$ plane, the allowed 
contours of global solar data and that of KamLAND data have a significant
overlap \cite{bahcall2004}. This overlap region is essentially what is obtained
if an analysis of all solar neutrino and KamLAND data is performed
\cite{abhi,deholanda,strumia,fogli,creminelli}.
However, in this paper, we address the following question:
Given $\Delta_{sol} = 8 \times 10^{-5}$ eV$^2$, what predictions do
we get for the solar neutrino spectrum? We first analyze the neutrino 
survival probability ($P_{ee}$) of the Large Mixing Angle (LMA) 
solution of the solar 
neutrino problem assuming two flavour oscillations. We study the 
energy dependence of $P_{ee}$ in detail as a function of the 
mixing angle $\theta_{sol}$ and show how these details can be tested 
in the future by precise data from Super-K and SNO experiments. We 
then demonstrate that a potential conflict with future spectral 
measurements can be resolved by reanalyzing $P_{ee}$ in a three 
flavour framework and by invoking a moderate value of $\theta_{13}$. 
Varying $\Delta_{sol}$ within the range allowed by 
KamLAND, does not change our conclusions.

\section{Two flavour analysis}

In this section we assume that only two mass eigenstates participate
in the oscillations of electron neutrinos and those of their anti-neutrinos.
This is equivalent to setting $\theta_{13} = 0$ in three flavour 
oscillations \cite{kuopan89} and we have 
$$
\Delta_{sol} = \Delta_{21} \ {\rm and} \ \theta_{sol} = \theta_{12}.
$$ 
We also assume CPT invariance and take the mass-square difference
measured by KamLAND to be $\Delta_{sol}$.
The electron neutrino survival probability, in the case of LMA solution
of the solar neutrino problem, is given by \cite{parke}
\begin{equation}
P_{ee} = \frac{1 + \cos 2 \theta^c_{12} \cos 2 \theta_{12}}{2},
\label{2flpee}
\end{equation}
where $\theta^c_{12}$ is the matter dependent value of $\theta_{12}$ 
at the solar core and is given by,
\begin{equation}
\cos 2 \theta^c_{12} = \frac{\Delta_{21} \cos 2 \theta_{12} - A^c}{
\Delta^c_{21}}. \label{th12c}
\end{equation}
Here $A^c ({\rm in~eV^2})= 10^{-5}~E \ ({\rm in~MeV})$ is the 
Wolfenstein term \cite{wolf} in the core of the sun and $\Delta^c_{21}$ 
is the corresponding mass-squared difference. It is given by
\begin{equation}
\Delta^c_{21} = \sqrt{(\Delta_{21} \cos 2 \theta_{12} - A^c)^2 +
(\Delta_{21} \sin 2 \theta_{12})^2}. \label{d21c}
\end{equation}
Analysis of solar neutrino data \cite{bahcall2004} restricts the 
$3 \sigma$ range of $\theta_{12}$ to be $27^\circ - 41^\circ$,
with the best fit at $34^\circ$.

The data from various gallium experiments \cite{gallium1,gallium2,gallium3}
requires that the average $P_{ee}$ for low energy ($E < 1$ MeV) 
should be about $0.54$ and Super-K spectrum data requires that 
for high energy  ($E > 5$) MeV, $P_{ee}$ should be essentially
constant with a value of about $0.35$ \cite{sk2002plb}. 
To see the variation of $P_{ee}$ with respect to energy, we
differentiate it and obtain 
\begin{eqnarray}
\frac{d P_{ee}}{dE} 
\ ({\rm per~MeV}) 
& = & \frac{\cos 2 \theta_{12}}{2}
\sin^2 2 \theta^c_{12} \frac{10^{-5}}{\Delta_{21}^c} \\
 & = & \frac{10^{-5}}{2 \Delta_{21}} 
\frac{\cos 2 \theta_{12} \sin^2 2 \theta_{12}}{
[(\cos 2 \theta_{12} - A^c/\Delta_{21})^2 + 
(\sin 2 \theta_{12})^2]^{3/2}}. 
\label{dpeede}
\end{eqnarray}
We now point out the two essential features of Eq.~(\ref{dpeede}).
\begin{itemize}
\item The slope of $P_{ee}$ is steepest at the energy when 
$A^c \approx \Delta_{21} \cos 2 \theta_{12}$. This energy of
steepest slope is given by $E_{ss} \approx 8 \cos 2 \theta_{12}$, 
where we have used the value of $\Delta_{21}$ from KamLAND.
\item In the neighbourhood of $E_{ss}$, the slope achieves its 
maximum value of $\cos 2 \theta_{12}/(16 \sin 2 \theta_{12})$. 
This shows that smaller the value of the vacuum mixing angle 
greater will be the slope of $P_{ee}$ in the neighbourhood of 
$E_{ss}$. 
\end{itemize}

\begin{figure}[hbt]
\begin{center}
  \includegraphics[width=0.83\textwidth]{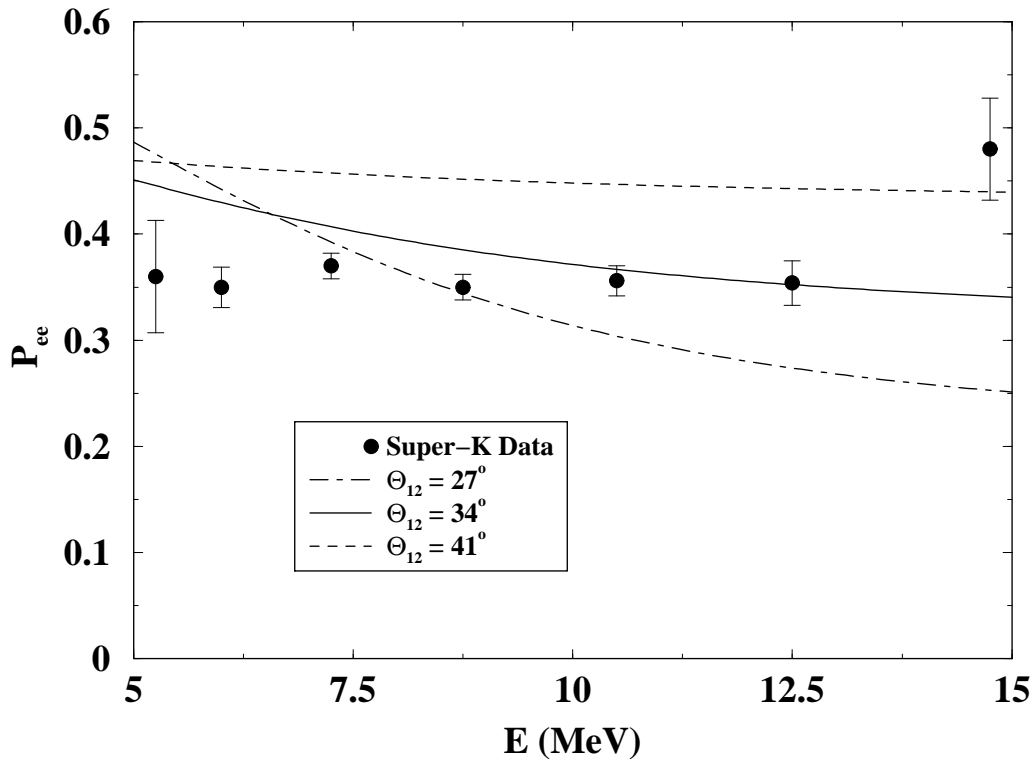}
\caption{$P_{ee}$ vs $E$ for the energy range 5-15 MeV
for $\Delta_{21} = 8 \times 10^{-5}$ eV$^2$ and for 
$\theta_{12} = 27^\circ, 34^\circ$ and $41^\circ$.
Data points with $1 \sigma$ error bars are obtained using the
Super-K data from \cite{sk2002plb}.}
\end{center}
\end{figure}

In figure~1, we have plotted $P_{ee}$ vs $E$, for three values 
of $\theta_{12}$, which are the best fit value and the lower
and the upper $3 \sigma$ bounds obtained by global analysis. 
As stated earlier, the value of $\Delta_{21}$ is fixed to be 
the KamLAND best fit value $8 \times 10^{-5}$ eV$^2$. 
To facilitate the comparison between LMA predictions and the 
data, the spectrum data from Super-K \cite{sk2002plb} are also 
shown in the figure. 

We observe the following important features of each of these plots.
\begin{enumerate}
\item
For a low value of $\theta_{12} = 27^\circ$, $P_{ee}$ decreases 
sharply from $0.49$ at $5$ MeV to $0.36$ at $8$ MeV and further 
decreases to $0.26$ at $14$ MeV. This is in conflict with the 
essentially flat spectrum observed by Super-K and SNO for $E > 
5$ MeV \cite{sk2002plb,snospect}. For this value of $\theta_{12}$,  
the energy region of the steepest slope in $P_{ee}$ is centered 
around $E_{ss} \approx 5~{\rm MeV}$. Hence there is a sharp drop 
in the probability in this range. The average value of $P_{ee}$
for this curve is close to the average $P_{ee}$ measured by 
Super-K but the sharp drop in the $5-8$ MeV region means that
the curve misses the points at both low and at high energies.
\item
For a high value of $\theta_{12} = 41^\circ$, we do have a flat 
spectrum, but the average value of $P_{ee}$ is $0.45$, which is 
considerably higher than $0.35$ measured by Super-K and SNO 
\cite{sk2002plb,snospect}. For this value of $\theta_{12}$,  
the energy region of the steepest slope in $P_{ee}$ is centered 
around $E_{ss} \approx 1.1~{\rm MeV}$, so the energy range $5-8$ 
MeV is reasonably away from this point. The flat spectrum in this case 
is the consequence of $\theta_{12}$ being close to maximal mixing, i.e
$45^\circ$, in which case matter effects play no role and where 
$P_{ee}$ is predicted to be close to $0.5$ for the whole energy range. 
\item 
For the best fit value of $\theta_{12} = 34^\circ$, $P_{ee}$ changes
from $0.45$ at $5$ MeV to $0.4$ at $8$ MeV and further decreases to
$0.35$ at $14$ MeV. For this value of $\theta_{12}$, the energy region 
of the steepest slope in $P_{ee}$ is centered around $E_{ss} \approx 
3~{\rm MeV}$. The energy range $5-8$ MeV is a little away from this 
point but well within the region where $dP_{ee}/dE$ (c.f. 
Eq.~\ref{dpeede}) is not negligible. 
\end{enumerate}

Thus we find that low values of $\theta_{12}$ give the correct
average value of $P_{ee}$ in the energy range visible at Super-K
but predict a steep slope. High values of $\theta_{12}$ have flat
spectrum but the average value of $P_{ee}$ is much higher than 
that measured by Super-K. The best fit value is a presently 
acceptable compromise between the two different pulls on $\theta_{12}$
generated by the two different features of $P_{ee}$, {\it i.e.}
its average value and its slope. But even the best fit curve 
suffers from two objections when compared to the data.
\begin{enumerate}
\item The predicted values of $P_{ee}$ in the energy range 
$5-8$ MeV are $2 \ {\rm to} \ 3 \sigma$ too high compared 
to the data.
\item The predicted values show slope in the $5-8$ MeV range
which is not observed in the data.
\end{enumerate}
These features are also reflected in table~1, where the present
Super-K data \cite{sk2002plb} and the two flavour LMA predictions 
for $P_{ee}$ for the best fit value of $\theta_{12}$ are shown. 
If future data from Super-K and SNO reduce the error bars in these 
bins by a factor $2$, and if the central values of the measurements 
do not change, then it must be concluded that two flavour oscillations 
are inadequate to explain both KamLAND data and the solar neutrino 
spectrum.

\begin{table}[htb]
\begin{tabular}{|c|c|c|c|}
\hline
Energy Range (MeV) & (Obs/SSM)$_{SK}$ & $P_{ee} (expt)$ & LMA-$P_{ee}$ \\
& & & $\theta_{12} = 34^\circ$ \\ \hline
$5 - 5.5$ & $0.467 \pm 0.04 \pm 0.017$ & $0.36 \pm 0.05 \pm 0.018$ & 
0.446 \\ \hline
$5.5 - 6.5$ & $0.458 \pm 0.014 \pm 0.007$ & $0.35 \pm 0.017 \pm 0.008$ & 
0.429 \\ \hline
$6.5 - 8$ & $0.476 \pm 0.008 \pm 0.006$ & $0.37 \pm 0.01 \pm 0.007$ & 
0.407 \\ \hline
$8 - 9.5$ & $0.460 \pm 0.009 \pm 0.006$ & $0.35 \pm 0.01 \pm 0.007$ & 
0.385 \\ \hline
$9.5 - 11.5$ & $0.463 \pm 0.01 \pm 0.006$ & $0.356 \pm 0.012 \pm 0.007$ &
0.367 \\ \hline
$11.5 - 13.5$ & $0.462 \pm 0.017 \pm 0.006$ & $0.354 \pm 0.02 \pm 0.007$ & 
0.352 \\\hline
$13.5 - 16$ & $0.567 \pm 0.039 \pm 0.008$ & $0.48 \pm 0.047 \pm 0.01$ & 
0.342 \\ \hline
$16 - 20$ & $0.555 \pm 0.146 \pm 0.008$ & $0.466 \pm 0.175 \pm 0.01$ & 
0.332 \\ \hline
\end{tabular}
\caption{$P_{ee}$ extracted from Super-K spectral data vs the values
predicted by the LMA solution. All error bars are $1~{\rm \sigma}$.}
\end{table} 

\section{Three flavour oscillations}
 
We now consider three flavour oscillations with non-zero $\theta_{13}$.
We are concerned only with $P_{ee}$ and $P_{\bar{e} \bar{e}}$ which
are independent of the mixing angle $\theta_{23}$ and CP-phase $\delta$ 
\cite{kuopan89} but do depend on $\theta_{13}$. Because $\Delta_{31} \gg
\Delta_{21}$, oscillations due to $\Delta_{31}$ are averaged out at 
KamLAND and we have
\begin{equation}
P_{\bar{e} \bar{e}} = 1 - \frac{1}{2} \sin^2 2 \theta_{13} - \cos^4 \theta_{13}
\sin^2 2 \theta_{12} \sin^2 \left( 1.27 \Delta_{21} L / E \right).
\end{equation} 
For this expression of $P_{\bar{e} \bar{e}}$ also, the position 
of the kink in the positron spectrum depends only on $\Delta_{21}$
and is independent of $\theta_{12}$ and $\theta_{13}$. Hence the
value of $\Delta_{21}$ determined by the spectral distortion data
of KamLAND, even in the case of three flavour oscillations, is the
same as in the case of two flavour oscillations and has the same accuracy.

$P_{ee}$ for solar neutrinos for the LMA solution is given by 
\cite{kuopan89,mnru}
\begin{equation}
P_{ee} = \cos^2 \theta_{13} \cos^2 \theta_{13}^c \left(
\cos^2 \theta_{12} \cos^2 \theta_{12}^c + \sin^2 \theta_{12} 
\sin^2 \theta_{12}^c \right)
+ \sin^2 \theta_{13} \sin^2 \theta_{13}^c,
\label{pee3}
\end{equation}
where $\theta_{12}^c$ is given by Eq.~(\ref{th12c}) 
and Eq.~(\ref{d21c}), with the modification that $A^c$ in those
equations should be replaced by $A^c \cos^2 \theta_{13}$.
Matter dependence of $\theta_{13}$ is given by 
\begin{equation}
\sin \theta_{13}^c = \sin \theta_{13} \left[1 + \frac{A^c}{\Delta_{31}} 
\cos^2 \theta_{13} \right] ; 
\cos \theta_{13}^c = \cos \theta_{13} \left[1 - \frac{A^c}{\Delta_{31}} 
\sin^2 \theta_{13} \right]  \label{th13c}.
\end{equation}
Here $\Delta_{31} = \Delta_{atm} = 2 \times 10^{-3}~{\rm eV}^2 \gg A^c$ 
for the whole energy range of solar neutrinos. Hence in 
Eq.~(\ref{pee3}) $\theta_{13}^c$ can be replaced by $\theta_{13}$
and $P_{ee}$, of three flavour oscillations, simplifies to:
\begin{equation}
P_{ee} = \cos^4 \theta_{13} \left( \cos^2 \theta_{12} \cos^2 \theta_{12}^c 
+ \sin^2 \theta_{12} \sin^2 \theta_{12}^c \right) + \sin^4 \theta_{13}.
\label{pee3sim}
\end{equation}
Thus the solar neutrino survival probability in three flavour oscillations
depends only on one additional parameter $\theta_{13}$. 

Analysis of only solar neutrino data allows $\theta_{13} \leq 40^\circ$
\cite{fogli2000} and analysis of only atmospheric neutrino data
allows $\theta_{13} \leq 30^\circ$ \cite{jimprl99,learned,marrone}. 
It is the CHOOZ experiment, which in combination with the 
$\Delta_{31}$ measurement from atmospheric neutrino data which 
sets a stringent upper bound on $\theta_{13}$ \cite{nruchz}. 
Present lower bound on $\Delta_{31}$ from Super-K data is $1.5 
\times 10^{-3}$ eV$^2$ \cite{sk05}. For this small a value of 
$\Delta_{31}$, CHOOZ sets an upper bound of about $15^\circ$ 
\cite{chooz}. 

The potential conflict between the LMA predictions for 
$\Delta_{21} = 8 \times 10^{-5}$ eV$^2$ and the Super-K measurement
of the solar spectrum can be naturally resolved in a three flavour 
oscillation framework. In the three flavour case, the energy range
where $A^c \cos^2 \theta_{13} \approx \Delta_{21} \cos 2 \theta_{12}$ 
is centered around the energy $E_{ss} = 8 \cos 2 \theta_{12} / \cos^2 
\theta_{13}$. Thus, for a given $\theta_{12}$, the strongest distortion 
is shifted to slightly higher energies as compared to the two flavour case. 
For even moderate values of $\theta_{13}$, $\cos^4 \theta_{13}$ differs 
appreciably from $1$. Hence, in three flavour oscillations, both $P_{ee}$
and its slope become smaller, thus mitigating the two discrepancies between
the solar spectrum data and LMA predictions of two flavour oscillations  
\cite{srubasmir}. 

First we take a look at low energies, where $E \leq 1$ MeV. In this
energy range the matter term $A^c \ll \Delta_{21}$ and we have
$\theta_{12}^c \simeq \theta_{12}$. Then Eq.~(\ref{pee3sim})
simplifies to:
\begin{eqnarray}
\langle P_{ee}^l \rangle & = & \cos^4 \theta_{13} \left(
\cos^4 \theta_{12} + \sin^4 \theta_{12}  \right)
+ \sin^4 \theta_{13}  \nonumber \\
&=& \cos^4 \theta_{13}\left(1-\frac{1}{2} \sin^2 2 \theta_{12} \right)
+ \sin^4 \theta_{13} \nonumber \\
&=& 1- \frac{1}{2} \sin^2 2 \theta_{13} - \frac{1}{2} \sin^2 2 \theta_{12} 
\cos^4 \theta_{13} . \label{pee3low}
\end{eqnarray}
If we set $\theta_{13} = 0^{\circ}$, Eq.(\ref{pee3low}) gives us the usual
low energy form for the LMA solution. From the gallium experiments
we have $P_{ee}^l \simeq 0.54 \pm 0.045$. In pure two flavour oscillations,
this data pulls $\theta_{12}$ towards $\pi/4$. However, we see
from Eq.~(\ref{pee3low}) that a moderate value of $\theta_{13}$
allows $\theta_{12}$ to differ appreciably from $\pi/4$.

To study the behaviour at high energies it is convenient to recast
Eq.(\ref{pee3}) as,
\begin{equation}
\langle P_{ee} \rangle  =  \cos^4 \theta_{13} \left(\frac{1}{2}+
\frac{1}{2} \cos 2 \theta_{12}^c \cos 2 \theta_{12}  \right)
+ \sin^4 \theta_{13}.   
\label{pee3alt}
\end{equation}
There is no immediate simplification of the above formula because
for the solar neutrino energy range the Wolfenstein term is never
much larger than $\Delta_{21}$. Hence we use the full expression
and study its dependence on $\theta_{13}$ for the best fit value 
of $\theta_{12}$, by means of graphs of $P_{ee}~{\rm vs}~E$. 

\begin{figure}[hbt]
\begin{center}
 \includegraphics[width=0.83\textwidth]{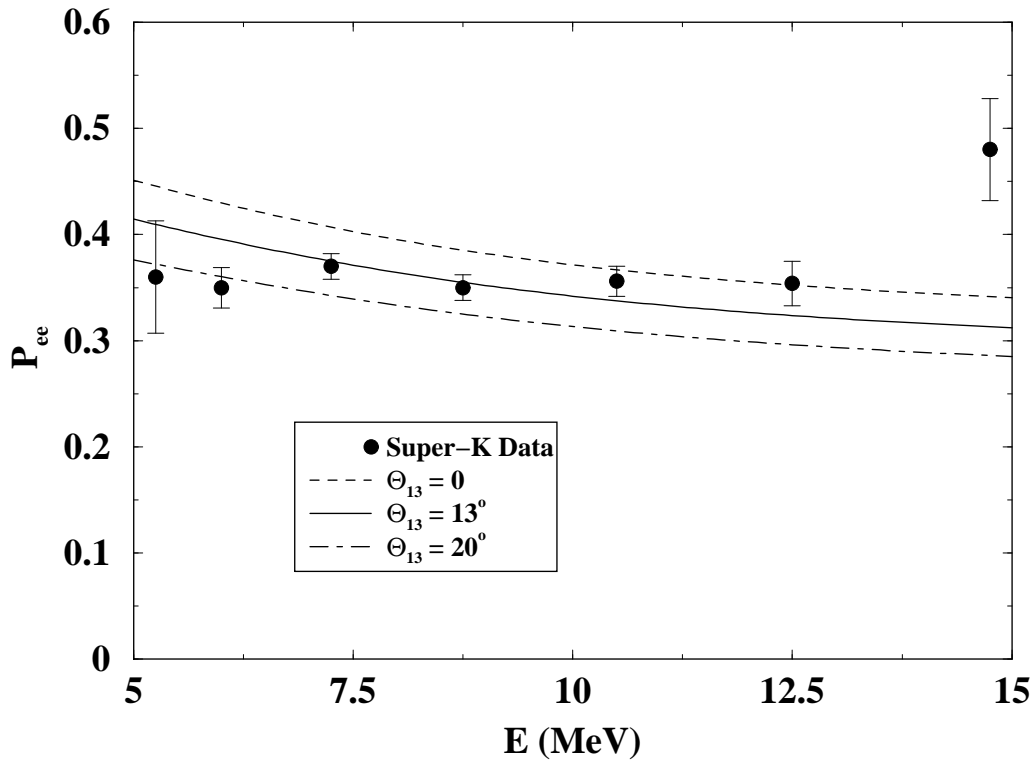}
\caption{$P_{ee}$ vs $E$ for the energy range 5-15 MeV
for $\Delta_{21} = 8 \times 10^{-5}$ eV$^2$,  
$\theta_{12} = 34^\circ$ and 
$\theta_{13} = 0^\circ, 13^\circ, 20^\circ$.
For comparison, Super-K data \cite{sk2002plb} with $1 \sigma$
error bars also are included.}
\end{center}
\end{figure}

In Fig.(2), for $\theta_{12} = 34^{\circ}$, we plot $P_{ee}$ 
for different values of $\theta_{13}$ along with the graph for 
$\theta_{13} = 0^{\circ}$.
Note that the effect of even moderate values of $\theta_{13}$ is to
pull the predictions for $P_{ee}$ towards the data, thus
leading to a better agreement.       

\section{conclusion}
Recent KamLAND data has unambiguously picked the parameters in the 
LMA region as the solution to the solar neutrino problem, confirming 
the expectations of global analyses of solar neutrino data.
However, for the large value of $\Delta_{21}$ obtained by KamLAND,
there is a discrepancy between the two flavour predictions of $P_{ee}$ 
and the Super-K spectrum data for all allowed values of $\theta_{12}$. 
For low values of $\theta_{12}$ the discrepancy is in the slope of 
$P_{ee}$ in the range $5-8$ MeV and for larger values of $\theta_{12}$
the discrepancy is in the average value of $P_{ee}$ in the energy 
range $5-15$ MeV. This discrepancy can turn into a full fledged conflict
if the error bars on the solar neutrino spectrum data can be reduced by 
a factor of 2. This conflict between the large value of $\Delta_{21}$ and 
the Super-K spectrum can be resolved in a three flavour oscillation analysis
for moderate values of $\theta_{13}$.  
A non zero value of $\theta_{13}$ of about $13^{\circ}$ can
both reduce the average value of $P_{ee}$ and give a flatter spectrum 
and the conflict with the spectrum measurement can be resolved. 
An equivalent and perhaps more striking way of looking at this analysis is 
that a careful study of solar neutrino spectrum gives us an
handle on $\theta_{13}$ {\it independent of the reactor constraints}. 

{\bf Acknowledgement} We thank Ameeya Bhagwat for his help in preparing
the figures.


\begin{references}
\bibitem{kamlandsp}
KamLAND Collaboration: T.~Araki {\it et al}, 
hep-ex/0406035.
\bibitem{srubabatikaml}
A.~Bandyopadhyay, S.~Choubey, S.~Goswami and S.~T.~Petcov,
hep-ph/0410283.
\bibitem{bahcall2004}
J.~N.~Bahcall, M.~C.~Gonzalez-Garcia, C.~Pena-Garay, JHEP {\bf 0408},
016 (2004), hep-ph/0406294.
\bibitem{abhi}
A.~Bandyopadhyay, S.~Choubey, S.~Goswami, S.~T.~Petcov, D.~P.~Roy,
hep-ph/0406328.
\bibitem{deholanda}
P.~C.~de Holanda and A.~Yu.~Smirnov, Astropart. Phys {\bf 21},
287 (2004), hep-ph/0212270.
\bibitem{strumia}
F.~Feruglio, A.~Strumia and F.~Vissani, Nucl. Phys. {\bf B 637},
345 (2002), (hep-ph/0201291) and {\it ibid} {\bf 659}, 359 (2003) (addendum).
\bibitem{fogli}
G.~L.~Fogli, E.~Lisi, A.~Marrone, D.~Montanino, A.~Palazzo and
A.~M.~Rotunno, Phys. Rev. {\bf D 67}, 073002 (2003), hep-ph/0212127 and
{\it ibid} {\bf 69}, 017301 (2004) (addendum), hep-ph/0308055.
\bibitem{creminelli}
P.~Creminelli, G.~Signorelli and A.~Strumia, JHEP {\bf 0105}, 
052 (2001) and updates of hep-ph/0102234.
\bibitem{kuopan89}
T.~K.~Kuo and J.~T.~Pantaleone, Rev. Mod. Phys. {\bf 61},  
937 (1989).
\bibitem{parke}
S.~J.~Parke, Phys. Rev. Lett. {\bf 57}, 1275 (1986). 
\bibitem{wolf}
L.~Wolfenstein, Phys. Rev. {\bf D 17}, 2369 (1978).
\bibitem{gallium1}
W.~Hampel {\it et al}, Phys. Lett. {\bf B 447}, 127 (1999).
\bibitem{gallium2}
C.~M.~Cattadori, Nucl. Phys. {\bf B 110}, Proc. Suppl, 311 (2002).
\bibitem{gallium3}
J.~N.~Abdurashitov {\it et al}, J. Exp. Theor. Phys, {\bf 95}, 181 (2002).
\bibitem{sk2002plb}
S.~Fukuda {\it et al}, Phys. Lett. {\bf B 539}, 179 (2002).
\bibitem{snospect}
The SNO collaboration: B.~Aharmim {\it et al}, nucl-ex/0502021.
\bibitem{mnru}
M.~Narayan, M.~V.~N.~Murthy, G.~Rajasekaran and S.~Uma Sankar
Phys. Rev. {\bf D 53}, 2809 (1996), hep-ph/9505281.
\bibitem{fogli2000}
G.~L.~Fogli, E.~Lisi, D.~Montanino and A.~Palazzo, Phys. Rev. {\bf D 62},
013002 (2000), hep-ph/9912231.
\bibitem{jimprl99}
J.~Pantaleone, Phys. Rev. Lett. {\bf 81}, 5060 (1998), hep-ph/9810467.
\bibitem{learned}
J.~Learned, hep-ex/0007056.
\bibitem{marrone}
G.~L.~Fogli, E.~Lisi, A.~Marrone, and D.~Montanino, Nucl. Phys. Proc.
Suppl. {\bf 91}, 167 (2000), hep-ph/0009269.
\bibitem{nruchz}
M.~Narayan, G.~Rajasekaran and S.~Uma Sankar
Phys. Rev. {\bf D 58}, R031301 (1998), hep-ph/9712409.
\bibitem{sk05}
Super-Kamiokande Collaboration: Y.~Ashie {\it et al}, hep-ex/0501064.
\bibitem{chooz}
CHOOZ Collaboration, Phys. Lett. {\bf B 466}, 415 (1999).
\bibitem{srubasmir}
S. Goswami and A. Smirnov, hep-ph/0411359.
\end{references}
\end{document}